\documentclass[twocolumn,showpacs,aps,prl,superscriptaddress]{revtex4}

\usepackage{graphicx}
\usepackage{verbatim}
\usepackage{dcolumn}
\usepackage{multirow}
\usepackage{bm}
\usepackage{amsmath}
\usepackage{amssymb}
\usepackage{epic}
\usepackage{overpic}
\usepackage{subfigure}

\input babarsym

\def\mrec{\ensuremath{M_{\mathrm{rec}}}\xspace}
\def\hephoton{\ensuremath{\gamma_{1}}\xspace}
\def\invisible{\ensuremath{\mathrm{invisible}}\xspace}
\def\MVA{\ensuremath{\mathrm{MVA}}}

\def\bfinvis{\ensuremath{\BR(\Y1S \to \invisible)}\xspace}

\newcommand{\ie}        {\mbox{\textsl{i.e.}}\xspace}
\newcommand{\vs}        {\mbox{\textsl{vs.}}\xspace}
\newcommand{\etal}      {\mbox{\textsl{et al.}}\xspace}

\newcommand{\BABARPubYear}    {09}
\newcommand{\BABARPubNumber}  {022}
\newcommand{\SLACPubNumber} {13757}

\def\figurebox#1#2#3{%
    \def\arg{#3}%
    \ifx\arg\empty
    {\hfill\vbox{\hsize#2\hrule\hbox to #2{\vrule\hfill\vbox to #1{\hsize#2\vfill}\vrule}\hrule}\hfill}%
    \else
    {\hfill\epsfbox{#3}\hfill}%
    \fi}

\begin{document}

\preprint{\babar-PUB-\BABARPubYear/\BABARPubNumber} 
\preprint{SLAC-PUB-\SLACPubNumber} 

\begin{flushleft}
\babar-PUB-\BABARPubYear/\BABARPubNumber\\
SLAC-PUB-\SLACPubNumber\\[10mm]
\end{flushleft}

\title{
{\Large \bf A Search for Invisible Decays of the $\mathbf{\Y1S}$}
}

%
\author{B.~Aubert}
\author{Y.~Karyotakis}
\author{J.~P.~Lees}
\author{V.~Poireau}
\author{E.~Prencipe}
\author{X.~Prudent}
\author{V.~Tisserand}
\affiliation{Laboratoire d'Annecy-le-Vieux de Physique des Particules (LAPP), Universit\'e de Savoie, CNRS/IN2P3,  F-74941 Annecy-Le-Vieux, France}
\author{J.~Garra~Tico}
\author{E.~Grauges}
\affiliation{Universitat de Barcelona, Facultat de Fisica, Departament ECM, E-08028 Barcelona, Spain }
\author{M.~Martinelli$^{ab}$}
\author{A.~Palano$^{ab}$ }
\author{M.~Pappagallo$^{ab}$ }
\affiliation{INFN Sezione di Bari$^{a}$; Dipartimento di Fisica, Universit\`a di Bari$^{b}$, I-70126 Bari, Italy }
% ...
\collaboration{The \babar\ Collaboration}
\noaffiliation

\date{\today}

\begin{abstract}
We search for invisible decays of the \Y1S meson
using a sample of $91.4 \times 10^{6}$ \Y3S mesons collected at the \babar/\pep2 
\BF.  We select events containing the decay $\Y3S \to \pipi \Y1S$ and search for evidence
of an undetectable \Y1S\ decay recoiling against the dipion system.  
We set an upper limit on the branching fraction 
$\bfinvis < 3.0 \times 10^{-4}$ at the 90\% confidence level.
\end{abstract}

\pacs{13.25.Hw, 12.15.Hh, 11.30.Er}

\maketitle

The nature of dark matter is one of the most challenging issues facing
physics. Observation of standard model (SM) particles coupling
to undetectable (invisible) final states might provide information on
candidate dark matter constituents. 
In the SM, invisible decays of the \Y1S meson proceed by \bbbar annihilation 
into a \nunub pair, with a branching fraction 
$\bfinvis \approx 1 \times 10^{-5}$~\cite{ref:Chang1997tq}, well below
the current experimental sensitivity.   However, 
low-mass dark matter candidates could couple weakly to SM 
particles to enhance the invisible branching fraction to the level of 
$10^{-4}$ to $10^{-3}$~\cite{ref:McElrath2005}.

Searches for this decay of the \Y1S can be carried out at \epem colliders
operating at the \Y2S or \Y3S resonance.  The presence of the \Y1S is
tagged by detecting the particles emitted in decays of the resonance to \Y1S.
Previous searches by the CLEO~\cite{ref:Rubin2006gc} 
and Belle~\cite{ref:Tajima2006nc} collaborations yielded upper limits of
$\bfinvis < 3.9\times 10^{-3}$ and $<2.5 \times 10^{-3}$ at the 90\%
confidence level (C.L.), respectively. In this paper we present a search 
for this final state using almost an order-of-magnitude more \Y1S mesons.

The data used in this analysis were collected with the \babar\ detector 
at the \pep2\ asymmetric-energy \epem collider 
running at an \epem center-of-mass (CM) energy corresponding to 
the mass of the \Y3S (10.3552\gevcc~\cite{ref:pdg2008}).
The presence of a \Y1S meson is tagged by reconstructing the
\pipi pair (\mbox{dipion}) in the transition $\Y3S \to \pipi \Y1S$.
The \babar\ detector is described in detail elsewhere~\cite{ref:babar}. 
These data were taken using an upgraded 
muon system, instrumented with both resistive plate chambers~\cite{ref:babar}
and limited streamer tubes between steel absorbers \cite{ref:lst_menges}. 
For these data the trigger was modified to substantially 
increase the dipion trigger efficiency for the signal process.
The data sample containing these improvements represents
$96.5 \times 10^{6}$ \Y3S mesons.

We model both generic \Y3S decays and the signal process using a Monte Carlo 
(MC) simulation based on \textsc{Geant4}~\cite{ref:geant}.
The $\Y3S \to \pipi \Y1S$ events are generated according to the matrix
elements measured by the CLEO collaboration~\cite{ref:Cronin-Hennessy2007}.
In signal events, the mass recoiling against the dipion (\mrec)
peaks at the \Y1S mass ($9.4603\gevcc$~\cite{ref:pdg2008}).  
The same is true for background 
events in which a real $\Y3S\to\pipi\Y1S$ transition occurs but the 
\Y1S final-state particles are undetected (``peaking background''). However,
the dominant background containing a pair of low-momentum pions does not
exhibit this structure (``non-peaking background'').
The analysis strategy is as follows: first apply selection criteria to 
suppress background, primarily the non-peaking component;
then fit the resulting \mrec spectrum to measure the peaking
component (signal plus peaking background).


We define three subsamples for both data and MC events.
The first of these, the ``invisible'' subsample, is designed to contain
signal events.  Events in this subsample have only two charged pions.
The other two subsamples are used to check and correct MC 
predictions for the processes which contribute to the peaking background 
in the invisible subsample.
The ``four-track'' subsample consists of events with two pions 
plus two tracks with high momenta in the CM frame, 
consistent with two-body decay
of the \Y1S where both final-state particles are detected. 
The ``three-track'' subsample consists of events containing two pions 
and only one high-momentum track, consistent with two-body \Y1S decay
where only one of the final-state particles is detected.  

We select events in the invisible subsample by requiring that there are 
exactly two tracks originating from
the interaction point (``IP tracks''), with opposite electric charge.  An
IP track is required to have a point of closest approach to
the interaction point within 1.5\cm\ in the plane transverse to the
beams and within 2.5\cm\ along the $z$-axis.  We further require these
tracks to each have CM-frame momentum $p^*<0.8\gevc$,
consistent with pions from the dipion transition.  The dipion system
is required to have an invariant mass satisfying $M_{\pi \pi} \in
[0.25,0.95] \gevcc$, compatible with kinematic boundaries 
($M_{\pi \pi} \in [2 M_{\pi},(M_{\Y3S}-M_{\Y1S})]$)
after allowing for resolution effects.  The dipion recoil mass is
\begin{equation}
\mrec^2 = s + M^2_{\pi \pi} - 2 \sqrt{s} E^*_{\pi \pi},
\end{equation}
where $E^*_{\pi\pi}$ is the CM energy of the dipion system and 
$\sqrt{s} = 10.3552\gevcc$. 
We require that $\mrec \in [9.41,9.52] \gevcc$.  
The efficiency of this selection for signal events is about 64\%,
due to the requirement of reconstructing the two pions.
All selection criteria were finalized without looking at data in a narrower 
\mrec ``signal region''  that, according to simulation, contains more than
99\% of the signal (see discussion of signal shape below for the precise
signal-region definition).

We select three-track and four-track events using the same dipion selection 
as in the invisible subsample.
We search for high-momentum tracks from the \Y1S decay 
(\ie, from $\Y1S\to\epem $ or $\Y1S\to\mumu$).  
We require that there be one or two additional IP tracks, 
each with $p^*>2.0\gevc$.  
If either of these tracks passes electron identification criteria, both
are treated as electrons; otherwise, both are treated as muons.  In
the former case we account for possible radiative energy loss due to
bremsstrahlung by pairing an electron with a photon emitted close in
angle and increasing the electron's energy and momentum by the energy
of this photon.
When two high-momentum tracks are present, we require that they have opposite
charge and a two-track invariant mass $\in [9.00,9.80] \gevcc$. We
remove photon conversions from these events by rejecting the event
if either pion satisfies electron-identification criteria.
This introduces a negligible efficiency loss: the probability of a pion to 
be misidentified as an electron is $\approx 0.1\%$.
Finally, we require that the mass difference between the $\pipi\ellell$ and
\ellell systems $\in [0.89,0.92] \gevcc$.

At this stage, the background level in the invisible subsample is several 
orders of magnitude larger than any hypothetical signal. We reject most
of this remaining background with a multivariate analysis (MVA), 
implemented as a random forest of decision
trees~\cite{ref:Breiman2001}.
The random forest algorithm is trained on signal MC events and 
5.3\% of data outside the signal region in \mrec.  
The contribution of peaking components to these data is negligible.
The data and signal MC events 
used to train the MVA are excluded from the rest of the analysis, leaving
$91.4 \times 10^{6}$ \Y3S events for use in the final result.

We use the following variables, which have been determined
to be only weakly correlated with \mrec, as inputs to the MVA: the probability 
that the pions originate from a common vertex; the laboratory polar angle
and transverse
momentum of the dipion system; the total number of charged tracks, IP tracks
or otherwise, reconstructed in the event; booleans that indicate whether
either pion has passed electron, kaon, or muon identification criteria; the
cosine of the angle (in the CM frame) between the highest-energy photon (\hephoton)
and the normal to the decay plane of the dipion system; the energy in the laboratory
frame of the \hephoton; the total neutral energy in the CM frame; and
the multiplicity of \KL candidates, defined using the 
shape and magnitude of the shower resulting from interactions in the calorimeter.

The selection on the MVA output is optimized by choosing the threshold
that achieves the minimum statistical uncertainty (dominated by background)
on \bfinvis and, 
in the null signal hypothesis,  the lowest upper limit at the 90\%~C.L.
Both were achieved by requiring an MVA output $> 0.8$ (where the full range is
0 to 1).  The efficiency of this criterion for signal-MC events
is 37.0\%, as compared to 0.8\% for data events outside the signal region.
The total efficiency of all trigger and event 
selection requirements is determined from signal-MC simulation to be 16.4\%.

\begin{figure}[htp]
 \centering
 \includegraphics[width=\linewidth]{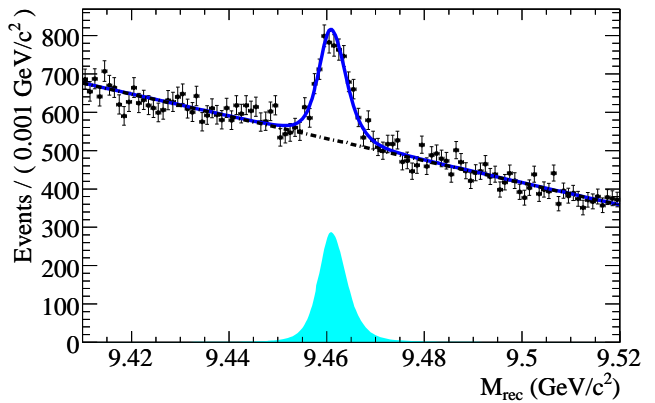}
 \vspace*{-0.3in}
 \caption{\label{fig:fit}
  The maximum likelihood fit to the dipion recoil mass 
  for data in the invisible subsample.  The components 
  of the fit are the non-peaking background (dash-dotted line) and 
  the peaking  component (solid filled).
  The total fit function is also shown (solid line).  
  Comparing the fit to the binned data, 
  $\chisq/\mathrm{(degree\ of\ freedom)} = 0.973$.
 }
\end{figure}

Figure 1 shows the \mrec distribution for the selected events.  We extract 
the peaking yield by an extended unbinned maximum likelihood fit, with
the non-peaking background described by a first-order polynomial.
The signal and the peaking background should have the same shape in
\mrec.  We describe this shape using a modified Gaussian function with a 
common peak position ($\mu_0$), independent left and right widths 
($\sigma_{L,R}$), and non-Gaussian tails (governed by parameters 
$\alpha_{L,R}$). The functional form on either side of the peak is 
\begin{eqnarray}
 \lefteqn{f_{L,R}(\mrec)= } \notag \\
 & \exp[-(\mrec-\mu_0)^2/(2\sigma_{L,R}^2+\alpha_{L,R}(\mrec-\mu_0)^2)]\ . &
\end{eqnarray}
We determine the parameters of this probability density function (PDF)
by fitting \mrec in the four-track data subsample.
The signal region, excluded when training the MVA, is defined as the region 
in Fig.~\ref{fig:fit} which is $<5\sigma_{L,R}$ from the peak
position, $\mrec \in [9.4487,9.4765] \gevcc$.  

The fit to the invisible subsample then determines
all of the parameters of the non-peaking background PDF, the
yield of the non-peaking background, and the yield of the peaking component.
The result for the peaking yield is $2326 \pm 105$ events.

Using a second-order polynomial for the non-peaking background results
in no change in the extracted peaking yield.
The systematic uncertainty on that yield associated with the fixed
parameters in the signal PDF is estimated by varying those parameters in the
fit.  We find an uncertainty of 18 events.  

We next estimate the contribution of background to the peak. 
The MC simulation predicts 1019 $\Y1S \to \epem$ events,  
1007 $\Y1S \to \mumu$ events,  
92 $\Y1S \to \tau^{+} \tau^{-}$ events, 
and $2.9 \pm 1.3$ $\Y1S \to \mathrm{hadrons}$ events.
These predictions depend upon branching fractions which
have significant uncertainties~\cite{ref:pdg2008}
and on the accuracy of the modeling of
event reconstruction and selection.  We use four-track and three-track 
data and MC subsamples
to test and correct the MC prediction of 2122 total events.

\begin{figure}[htp]
 \centering
 \includegraphics[width=0.49\linewidth]{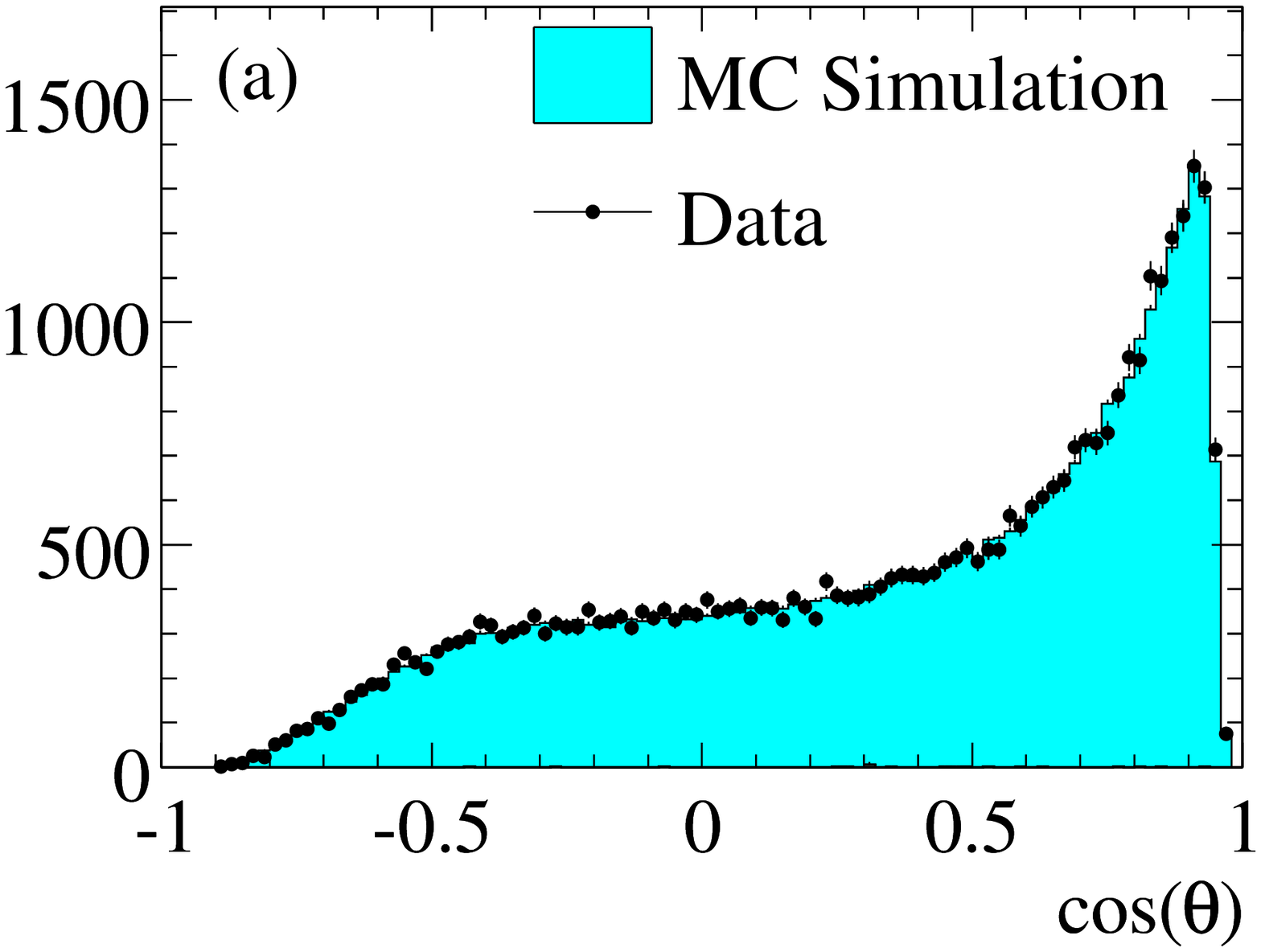}
 \includegraphics[width=0.49\linewidth]{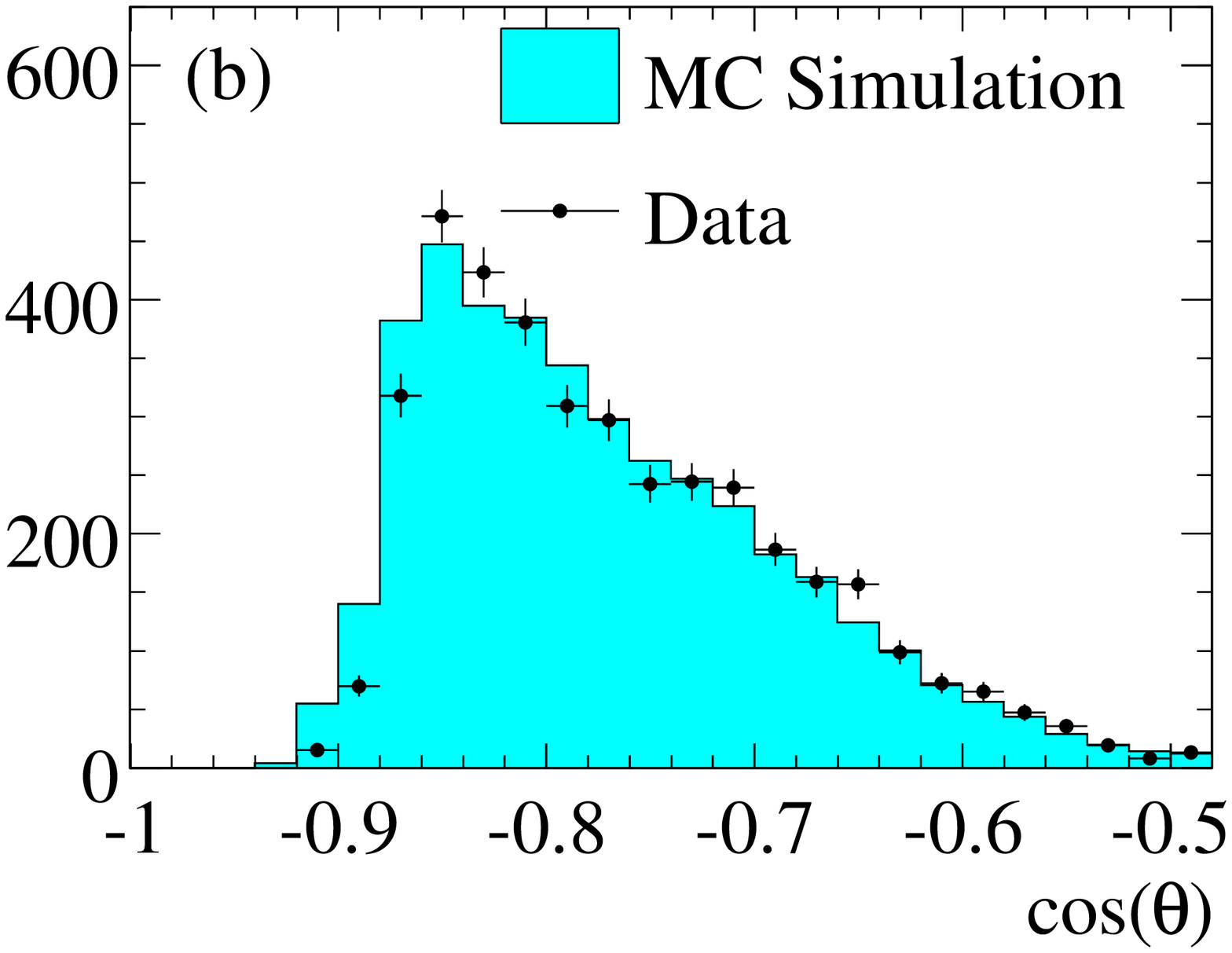}
 \caption{\label{fig:peaking}The distribution of $\cos(\theta)$, 
 where $\theta$ is the laboratory-frame polar angle of
 (a) the positively charged high-momentum track in the four-track subsample
 and (b) the single high-momentum track in the three-track
 subsample.  The normalization correction from the four-track 
 subsample has been applied to the MC yields in both cases.}
\end{figure}

We first use the four-track subsamples to calibrate the product of the
branching fractions for $\Y1S\to\ellell$ and $\Y3S\to\pipi\Y1S$ and
the dipion reconstruction efficiency.
We compare the event yields between four-track data and MC subsamples
when the positively-charged lepton is emitted in the central
section of the tracking system, $|\cos(\theta_{l^{+}})| < 0.3$
(laboratory-frame angle).  The simulation underestimates the number of events in data
by a factor of $(1.088 \pm 0.012)$.  This is plausible in light of
the branching fraction uncertainties and 
track reconstruction uncertainties ($\approx 0.5\%$ per track). 
Since the effect of the high-momentum track
reconstruction has a negligible contribution here, this data/MC correction 
factor is applied to \emph{all} of our MC-simulation subsamples.
For the four-track subsample, Fig.~\ref{fig:peaking}(a) shows that
the distribution of the high-momentum tracks in the detector is
very well described by the MC simulation at all polar angles.

We next compare the data and MC 
efficiencies for reconstructing the single lepton in the
three-track subsample.  Any discrepancy would imply a complementary
mistake in the invisible peaking background.
Given the CM-frame polar angle coverage of the detector, for three-track
events the high-momentum lepton in the forward direction often
escapes detection and thus the detected lepton is in the
backward direction.  We compare the MC and data 
laboratory-frame polar angle distributions for these 
events in Fig.~\ref{fig:peaking}(b). 
The three-track subsample, in contrast to the four-track subsample, has a  
significant non-peaking background in recoil mass.  Hence 
three-track peaking yields
\vs polar angle are determined by using the \mrec fit described above and
applying an event-weighting technique~\cite{ref:sPlots}.
The MC simulation describes the distribution well everywhere except 
at $\cos(\theta_{\ell}) < -0.84$,
where the simulation overestimates the reconstruction rate.

For leptons in this far-backward region, we use the ratio of data to
simulation \vs lepton $\cos(\theta)$ from Fig.~\ref{fig:peaking}(b)
as the basis of an accept-reject method applied to the high-momentum
track.  When this method removes the track, it in effect reassigns a 
three-track event to the invisible category.  We also weight the 
reassigned events by the ratio of simulated trigger efficiencies for
the three-track and invisible subsamples and assign 100\% uncertainty to this
difference in trigger efficiency.  Applying this additive correction after the 
scaling correction (from the four-track subsample), the total
peaking background estimate increases from 2122 events to
($2451 \pm 38$) events.

We test the prediction of the contribution of non-leptonic \Y1S decays to
the peaking background using an additional control sample.  Events in
this sample contain only two tracks (the pions) and pass all other 
criteria for the invisible subsample, except that the MVA requirement is
replaced by a requirement that the \hephoton has energy $>0.250\gev$. This
selects a set of events which is almost orthogonal to the signal
selection, since the $\MVA>0.8$ requirement results in a steep falloff
in efficiency \vs \hephoton energy near 0.250\gev.  We compare this
energy distribution in data (using the weighting
technique~\cite{ref:sPlots}) to that from simulation.  
As the \hephoton energy approaches $0.250\gev$ from above, we find the MC 
simulation underestimates the data by no more 
than a factor of four. Since the expected contribution of these events 
to the peaking background is 0.14\% of the total, we assign 0.6\% (15
events) as an additional systematic uncertainty on the peaking background,
for a total of $\pm 41$ events.

A number of multiplicative systematic corrections and uncertainties to
the peaking background also enter, in a fully-correlated manner,
when the extracted signal yield is converted to the $\Y1S \to \invisible$ 
branching fraction.  
The first such contribution is the $1.088 \pm 0.012$ correction factor
derived from the four-track subsample.  But this does not account for
trigger and MVA effects which might differ for the
invisible and four-track subsamples.  Since events
used to train the MVA have already passed the trigger requirements,
we first study the effect of trigger selection on data.
The \babar\ trigger consists of a hardware and a software stage.
The latter is tested by using a heavily-prescaled sample of
events which bypassed it.  We apply the software-level trigger to these events
and find that the ratio of efficiencies in data and MC simulation is 
$0.997 \pm 0.009$.  This ratio is taken as 
a correction to the signal efficiency and the peaking background.
To assess how well the 
impact of the hardware trigger on the two pions is simulated,
four-track events are used, since their trigger decision is
based largely on the two high-momentum lepton tracks.
We apply to the pions a set of selection 
criteria which approximate those applied by the hardware trigger.  The
data and MC efficiencies for these requirements
differ by 2.2\%.  Since this test is done on samples for which the 
hardware trigger is only approximated, we take this difference as a 
systematic uncertainty rather than apply a correction for it.

After applying the approximate hardware trigger criteria
to the four-track subsamples for both \Y3S MC simulation 
and data, we apply the nominal MVA selection to both.  The relative 
difference in efficiency between these MC and data subsamples is
4.0\%.  Since the hardware trigger is again only approximated for this test, 
we apply no correction for the difference, but assign it as a systematic 
uncertainty on the MVA selection.

Adding the multiplicative uncertainties in quadrature, the total correlated 
systematic uncertainty is 4.8\%.  The final corrected prediction 
for the peaking background is ($2444 \pm 123$) events, including the
uncorrelated uncertainty of 41 events.
From this we determine the signal
yield to be $(-118 \pm 105 \pm 124)$ events, where the errors are
statistical and systematic, respectively.  To obtain \bfinvis we
divide this by the signal efficiency, the number of \Y3S mesons, the 
branching fraction for the dipion transition ($4.48\%$ \cite{ref:pdg2008})
and the correction factors ($1.088 \times 0.997$).
The factor derived from the four-track subsample includes a possible
adjustment of $\BR(\Y1S\to\epem) + \BR(\Y1S\to\mumu)$, not relevant for
signal.  We take this adjustment to be $1.000 \pm 0.025$~\cite{ref:pdg2008} and
remove it by assigning an additional systematic uncertainty of 2.5\%.
Taking correlations into account, we determine that 
$\bfinvis = (-1.6 \pm 1.4\stat \pm 1.6\syst) \times 10^{-4}$.

Lacking evidence for this decay, we use a Bayesian technique
to set an upper limit on the branching fraction. 
We convolute the statistical likelihood, a function of \bfinvis,
with Gaussian functions representing the systematic error.   We assume a 
prior probability that is flat in branching fraction and integrate the
likelihood from 0 to a value such that 90\% of the total integral 
above 0 is enclosed.  The resulting limit is 
$\bfinvis < 3.0 \times 10^{-4}$ at the 90\% C.L.

In conclusion, we search for invisible decays of the
\Y1S meson.  We do so by looking for evidence of the
decay of the \Y1S into undetectable final states recoiling
against the dipion system in $\Y3S\to\pipi\Y1S$,
using a sample of $91.4 \times 10^{6}$ \Y3S mesons. 
We find no evidence for $\Y1S\to\invisible$ and set an upper limit
on its branching fraction at $3.0 \times 10^{-4}$ at the 90\% C.L.
This limit is almost an order of magnitude closer to
the SM prediction than the best previous limit.


We are grateful for the excellent luminosity and machine conditions
provided by our \pep2\ colleagues, 
and for the substantial dedicated effort from
the computing organizations that support \babar.
The collaborating institutions wish to thank 
SLAC for its support and kind hospitality. 
This work is supported by
DOE
and NSF (USA),
NSERC (Canada),
CEA and
CNRS-IN2P3
(France),
BMBF and DFG
(Germany),
INFN (Italy),
FOM (The Netherlands),
NFR (Norway),
MES (Russia),
MEC (Spain), and
STFC (United Kingdom). 
Individuals have received support from the
Marie Curie EIF (European Union) and
the A.~P.~Sloan Foundation.

\end{document}